\documentclass[3p,onecolumn]{elsarticle}

\usepackage{lineno,hyperref}
\usepackage{xcolor}
\usepackage{booktabs} 
\usepackage{multirow}
\usepackage{caption}
\usepackage{subcaption}
\usepackage{array}
\newcolumntype{L}[1]{>{\raggedright\arraybackslash}m{#1}}

\journal{International Journal of Human-Computer Studies}

\DeclareUnicodeCharacter{0301}{}

\usepackage[nolist]{acronym}
\begin{acronym}
    \acro{PwD}{People with Dementia}
    \acro{QoL}{Quality of Life}
    \acro{LSB}{Life Story Book}
    \acro{CEFR}{Common European Framework of Reference for Languages}
    \acro{IRB}{Institutional Review Board}
\end{acronym}

\begin{document}

\begin{frontmatter}

\title{Developing Assistive Technology to Support Reminiscence Therapy:\\ A User-Centered Study to Identify Caregivers' Needs}

\author[LASIGE]{Soraia M. Alarc\~{a}o\corref{mycorrespondingauthor}}
\ead{smalarcao@ciencias.ulisboa.pt}
\cortext[mycorrespondingauthor]{Corresponding author at: Departamento de Inform\'{a}tica, FCUL, Edif{\'{\i}́}cio C6, Piso 3, Sala 6.3.35. 1749-016 Lisboa, Portugal}

\author[LASIGE]{Andr\'{e} Santana}
\ead{santana.afm@gmail.com}

\author[LEL]{Carolina Maruta}
\ead{carolmaruta@gmail.com}

\author[LASIGE]{Manuel J. Fonseca}
\ead{mjfonseca@ciencias.ulisboa.pt}

\address[LASIGE]{LASIGE, Faculdade de Ci\^{e}ncias, Universidade de Lisboa, Portugal}
\address[LEL]{Laborat\'{o}rio de Estudos de Linguagem, Centro de Estudos Egas Moniz, Faculdade de Medicina, Universidade de Lisboa, Portugal}

\begin{abstract}
Dementia represents a major cause of disability among elderly subjects worldwide.
Reminiscence therapy is an inexpensive non-pharmacological therapy commonly used due to its therapeutic value for \ac{PwD}, as it can be used to promote independence, positive moods and behavior, and improve their quality of life.
Caregivers are one of the main pillars in the adoption of digital technologies for reminiscence therapy, as they are responsible for its administration.
Despite their comprehensive understanding of the needs and difficulties associated with the therapy, their perspective has not been fully taken into account in the development of existing technological solutions.
To inform the design of technological solutions within dementia care, we followed a user-centered design approach through worldwide surveys, follow-up semi-structured interviews, and focus groups. 
Seven hundred and seven informal and 52 formal caregivers participated in our study.
Our findings show that technological solutions must provide mechanisms to carry out the therapy in a simple way, reducing the amount of work for caregivers when preparing and conducting therapy sessions.
They should also diversify and personalize the current session (and following ones) based on both the biographical information of the \ac{PwD} and their emotional reactions.
This is particularly important since the \ac{PwD} often become agitated, aggressive or angry, and caregivers might not know how to properly deal with this situation (in particular, the informal ones).
Additionally, formal caregivers need an easy way to manage information of the different \ac{PwD} they take care of, and consult the history of sessions performed (in particular, to identify images that triggered negative emotional reactions, and consult any notes taken about them).
As a result, we present a list of validated functional requirements gathered for the \ac{PwD} and both formal and informal caregivers, as well as the corresponding expected primary and secondary outcomes.
\end{abstract}

\end{frontmatter}

\section{Introduction}

    Dementia represents a major cause of disability among elderly subjects worldwide \cite{Who2017,Prince2015}.
	The need for improvement of \ac{PwD}'s quality of life, independence, and positive behaviors, while reducing their families' burden has become a challenge in this field and motivated the development of intervention programs with positive outcomes \cite{Mulvenna2011,Kerssens2015,Dutzi2017}.
	
	Reminiscence is an inexpensive non-phar\-ma\-co\-log\-i\-cal therapy that is commonly used in dementia care due to its therapeutic value for elderly adults \cite{Cotelli2012,Jo2015,Siverova2014}, since it promotes communication between patients and the rest of the world.
	Older memories and events related to past information triggered by specific stimuli (e.g., photos, music, videos, books, newspapers, etc.) are recalled in sessions.

    Over the years, several computerized systems have been presented to help \ac{PwD} perform everyday activities, improve the therapeutic treatment of dementia by stimulating memories, communication, and social engagement, and finally to diagnose and evaluate progression of the disease \cite{Cohene2007,Dishman2007,Smith2009,Thiry2012b}.

    These solutions use a collection of multimedia data as stimuli that remains unchanged throughout therapy sessions.
    Although the possibility of customizing therapy sessions should be a must-have, the current solutions available may turn it into a burden for the caregivers  since they require a demanding choice of adequate multimedia contents.
	Moreover, existing solutions do not consider emotions related to the multimedia used in sessions nor the patient's emotional reactions to that multimedia.
	For example, a given image may trigger different emotions on different \ac{PwD}, which is particularly important since each person has their own memories associated with different stages of the disease.
	Ultimately, this could elicit negative feelings and unpleasant memories that informal caregivers may find hard to deal with \cite{Cotelli2012,Thiry2012}.
	
	Caregivers are one of the main pillars in the adoption of reminiscence therapy as they are responsible for its administration at home or in institutions.
    Thus, also responsible for adopting technology that may aid in therapy.
    However, and despite their comprehensive understanding of the needs and difficulties associated with reminiscence therapy, their perspective has not been fully taken into account in previous solutions.
    To inform the design of novel technological solutions within dementia care, we conducted an worldwide user-centered study with both formal and informal caregivers.
    We conducted three cross‐sectional, open‐access, web‐based surveys, followed by semi-structured interviews and focus groups.
    A total of 707 informal and 52 formal caregivers participated in the study.

    The current study aims to address the existing knowledge gap by investigating caregivers' perceptions in order to identify: i) whether reminiscence therapy is currently performed, and if not, what are the main reasons for that; ii) conditions in which sessions are performed (e.g. duration and frequency of sessions, materials used; iii) which emotional reactions emerge during sessions and how caregivers deal with them; iv) procedures related with the creation and use of a \ac{LSB}; v) considering the shortfalls identified in existing solutions, the caregivers' interest in technological functionalities regarding the management of \ac{PwD}, consultation of both the materials used in sessions and the history of sessions performed, management of multimedia material, automatic adaptation of the session's content considering the emotional information generated during sessions, and automatic gathering of new material personalized to each \ac{PwD}.

\section{Background and Related Work}

    In this section, we start by providing some background on dementia and reminiscence therapy, followed by the current state of the art regarding technological solutions to support reminiscence therapy.

    \subsection{Dementia and Reminiscence Therapy}
    
        Dementia is a clinical syndrome characterized by a significant cognitive decline in several cognitive (such as attention, executive function, learning and memory, language) and behavior domains, severe enough to interfere with personal, occupational and social activities, from a previous level of performance  \cite{NICE2006,APA2013}.
    
        The functional and behavioral dysfunction that comes with dementia is one of the leading causes of disability worldwide. 
        It has a significant impact on the lives of the \ac{PwD}, as well as in the quality of life of their families \cite{Cotelli2012}.
        For these reasons, dementia is associated with complex needs that often challenge the skills and capacity of caregivers and services, especially in the later stages, due to high levels of dependency and morbidity.
        
        According to the literature, psychological therapies slow the worsening of \ac{PwD}'s condition.
        Reminiscence therapy is the most universal non-pharmacological therapy used \cite{Cotelli2012,Jo2015,Siverova2014}.
        It is an effective treatment that promotes communication between \ac{PwD} and the rest of the world by stimulating their memories (from daily routines or their past). 
        It uses the preserved abilities rather than emphasizing the existing impairments, while helping to alleviate the experience of failure, social isolation, and decline of the well-being of \ac{PwD} \cite{Pringle2013}.
    
    \subsection{Technology for Reminiscence Therapy}

        Digital technology has become an indispensable tool that can be used to promote independence, positive moods and behavior, and improve the quality of life of \ac{PwD} without overburdening caregivers \cite{Kerssens2015, Mulvenna2011}.
        Over the last years, some computerized systems have been presented to help \ac{PwD} perform everyday activities, and complement the application of reminiscence therapy by stimulating memories, communication, and social engagement.
    	The majority of these solutions obtain information about the \ac{PwD} from multimedia biographies provided by family and friends \cite{Ibarra2017, OConnor2019}, while helping caregivers dialogue with patients based on their information and interests \cite{Webster2011}.
    
        CIRCA \cite{Gowans2004, Gowans2007}, Living in the moment \cite{Astell2009}, and CART \cite{Pringle2013} projects improve communication using videos and music, while the digital personal life history \cite{Cohene2007}, Companion \cite{Kerssens2011, Kerssens2015} and InspireD \cite{Laird2018} try to provide personalized, positive, and stimulating experiences using multimedia and trusted messages.
    	Collegamenti \cite{Ibarra2017} uses personal images, while MyStory and SharedMemories \cite{Edmeads2019} use audio and images to support meaningful interactions and collect stories that summarize important moments in the life of the \ac{PwD}.
    	DARE \cite{Gary2012} uses a multimedia album to help the \ac{PwD} reminisce, while identifying the image/video that generates a greater response in the brain.
    	Other approaches explore various sensory outputs (video, sound, objects, and smell) \cite{Mertl2019}.
    	
    	More recently, tangible and immersive solutions have emerged.
        The former includes a tangible multimedia book \cite{Huldtgren2016} allowing to touch, turn and press on pages; a chest of drawers \cite{Ly2016} where each drawer contains items representing sub-topics illustrated through images; and an old-style radio/TV \cite{Wilkinson2017} to present multimedia, enhancing the sense of familiarity and comfort.
        The latter group includes a companion robot \cite{Asprino2019} to deliver personalized interactive therapy through dialogue-based interactions complemented with multimedia; a multi-sensory stimulation solution \cite{Sorgaard2018}, combining images, sound, sights, smells, and movements to link the stimulation experience to the personal history of the \ac{PwD}; and the display of life size, hyper-real, software-rendered images \cite{Watson2018} that are navigable through space and time thus eliciting a sense of involvement.
    
        Most of these solutions use the same content throughout sessions, with few using information gathered to learn more about the people and to use this new knowledge in the future.
        In particular, the vast majority of the solutions does not assess how the \ac{PwD} reacts emotionally to these stimuli, nor it uses this information to automatically adapt the rest of the session or future therapy sessions.
        They also lack the use of personalized content that could increase the levels of enthusiasm and engagement of the \ac{PwD}. 
        All the solutions ignore the emotional content conveyed by the multimedia used.
        
        Participatory design is very important when designing any technological solution: it is important to understand the needs of the end users of such solutions, in this case, the caregivers of \ac{PwD} (that work as proxies of the \ac{PwD}), who will execute reminiscence therapy sessions (with the help of the solutions presented).
        The effort that authors of the majority of the solutions made in ensuring that they were evaluated by \ac{PwD} or dyads caregiver-PwD is noteworthy.
        However, requirements gathering (before developing the solutions) was only performed in about 40\% of them. 
        When it was done, it was usually with a small number of participants (exceptions are the solutions presented in \cite{Gowans2007,Ly2016,Edmeads2019}), in partnership with local institutions.
        As stated by Edmeads \textit{et. al}, it may lead to assumptions made about the issues and needs of \ac{PwD} and caregivers, and as result, about the technology that may be appropriate to address them. 
        As a major consequence, it might compromise the adoption and suitably of such technology for the users.
\section{User-Centered Study Methods}

\begin{figure}[!htp]
        \centering \includegraphics[width = 0.75\linewidth]{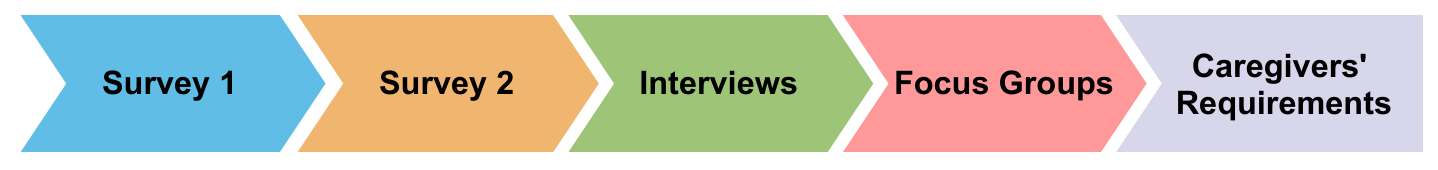}
    	 \caption{\label{fig:process} User-centered Approach for the Requirements Elicitation. (figure best seen in color)}
    \end{figure}

    We followed a user-centered design in which the results of each step of the study inform the next ones (see~Figure \ref{fig:process}).
    We started with a survey targeted at informal caregivers to identify the needs of both the \ac{PwD} and informal caregivers when performing reminiscence therapy (Survey 1).
    We further complement it with another survey aimed at both informal (to validate findings from Survey 1) and formal caregivers (to identify their needs in the context of therapy performed in health institutions) (Survey 2).
    Follow-up interviews and focus groups were conducted to gather more in-depth information, as well as to validate our main findings.

    Following, we detail how the selection of participants was carried out, the materials used, as well as the procedures adopted for the data collection and analysis of the results.

    \subsection{Participant Selection}
    
        Recruitment occurred via social and institutional networking.
    	Surveys were disseminated in Facebook\footnote{\url{https://facebook.com/}} groups selected according to the following keywords: \textit{alzheimer}, \textit{demência}, \textit{demens}, \textit{demenz}, \textit{demenza}, \textit{dementia}, \textit{huntington}, \textit{parkinson}, \textit{pick}, and \textit{vascular dementia}.
    	
    	We contacted several social/community/health-care institutions to disseminate the surveys through their communication channels. 
    	The process to select institutions was similar to the one used to select the Facebook groups.
    	
    	Caregivers who volunteered to participate in the interviews during the surveys were contacted and invited to participate.
        We also disseminated an invitation to participate in the interviews in the Facebook groups and social/\-community/\-health-care institutions where we had disseminated both surveys.

    \subsection{Materials}
    
        The first survey consisted of 65 open and closed-type questions tackling 4 domains: 1) socio-de\-mo\-graph\-ic characteristics (of both caregivers and \ac{PwD}); 2) reminiscence therapy (i.e. the process of revisiting older memories); 3) the use of an \ac{LSB}; 4) interest in technology.
    	Closed-type questions included Likert-type scales, ranging from 1 (few or very little) to 5 (too much) or 1 (does not do) to 4 (does it alone), multiple choice, and dichotomic items.
    	
    	Our second survey had two versions: one focused on informal caregivers, and another on the formal ones.
        The version for the informal caregivers consisted of 38 open/closed questions, while the one for the formal caregivers had 36 open/closed questions.
        Closed-type questions included Likert-type scales, ranging from 1 (few or very little) to 5 (too much), multiple choice, and dichotomic items.
        Both versions tackled the same domains: 1) socio-de\-mo\-graph\-ic characteristics of caregivers; 2) \ac{PwD} follow-up (reminiscence sessions, follow-up appointments, and other forms of follow-up); 3) interest in technology.
    	
    	We made our surveys available in three languages: English, Portuguese, and Spanish (only the first survey).
    	Portuguese and Spanish versions were proof-read by native speakers, while the English ones were proof-read by a subject with a C2 proficiency certification (according to the \ac{CEFR}).
    
        We conducted the semi-structured interviews and focus groups remotely, focusing on the characterization of the caregivers, their understating of what reminiscence therapy is, the care they provide, and which technology they are familiar with.
    
    \subsection{Procedures}
    
        Our study was designed in accordance with the Declaration of Helsinki, and national and legal regulatory requirements.
    	There were no anticipated risks involved in the participation.
    	The study protocol was submitted to the local \ac{IRB}, and was exempted from the full ethical application.
    	
        Surveys were accessible through a link to the Google Forms platform\footnote{\url{https://docs.google.com/forms/}} for each language and caregiver type.
        Interviews and focus groups were performed using the Zoom\footnote{\url{https://zoom.us/}} platform.
        Participation in each phase of the study was voluntary, and participants could withdraw at any point (since all the research was conducted online, participants had the option of not submitting the form in the case of the surveys or not finishing the interview/focus group).
    	Participants were not paid money nor received any other type of compensations, and all of them provided consent for the study.
    	
    	Data collection was anonymous, and no personally identifiable information was collected with the exception of e-mail (provided by the subject if they wished to be informed about the future work), which was only accessible to the investigators and kept confidential and off the records.
    
        Our user-centered study spans for about four and a half years.        
        The first survey data was collected during an 18-month interval, from July 2017 to December 2018.
        The second survey data was collected from March 2021 to July 2021.
        Follow-up interviews were conducted in July 2021, and from October to November 2021.
        Two focus groups were conducted on November 2021.
        Each interview lasted about 1 hour, and focus groups around 2 hours.
    
    \subsection{Data Analysis}
    
        Since the results drawn from each step of the study are used to inform the following ones, we needed to analyze each step individually.
        
        We conducted a mixed deductive and inductive thematic analysis on all the open-ended questions of the surveys, carried out in accordance with the strategy suggested by Braun and Clarke \cite{Braun2006}. 
        The content of open-questions was analysed several times and further categorized (I.M.).
    	Each response was coded by two independent investigators (I.M. and S.M.A.) according to the coding frame developed by I.M.
    	
    	The coding frame created in the first survey was used as a basis for the second one, having been adapted according to the characteristics of the second survey. 
        The content of open-questions from the second survey was analysed several times and further categorized (A.S.).
        Each response was coded by two independent investigators (A.S. and S.M.A.) based on the coding frame developed by A.S.
        For both surveys, two other investigators (M.J.F. and C.M.) acted as independent referees in case of divergence.
    	In general, a response could be coded for more than one category.
    	Demographics variables and categories were summarized by descriptive statistics.
    	  
    	Since all of our data involved nominal or ordinal variables, all statistical tests conducted were non-parametric: Spearman Rank, Chi-Square/Fisher's exact test, Kruskal-Wallis H, and Mann-Whitney U.
    	If Chi-Square/Fisher's exact test was significant, adjusted residuals with Bonferroni correction post hoc analysis was conducted.
    	If Mann-Whitney U test was significant, Kruskal-Wallis W was used as post hoc analysis.
        Finally, if Kruskal-Wallis W was significant, pairwise post-hoc Dunn test with Bonferroni adjustments was conducted.
    	Results of statistical tests were considered significant if the \textit{p}‐value was below 0.05.
    	Analyses were performed using IBM SPSS Statistics Version 26.
    	
    	Interviews and focus groups were recorded (with the permission of participants), and their audio recordings transcribed for analysis.
        We used verbatim transcription since it accounts for a richer interaction experience and is better for qualitative research \cite{Blake1995}.
        Each interview and focus group was coded by two independent investigators (A.S. and S.M.A). 
        A third investigator (M.J.F.) acted as an independent referee in case of divergence.
\section{Survey 1}
    
        In this section, we present our results regarding the sample characteristics (of both caregivers and \ac{PwD}), how reminiscence therapy is conducted, the creation and use of \acp{LSB}, and the interest in multiple technological functionalities (personalization of stimuli, identification of the \ac{PwD} emotional reactions to stimuli, and adaptation of current and future sessions).
    
        \subsection{Sample Characteristics}
            
            A total of 661 participants from 39 countries worldwide answered the survey.
    	    Fifty-eight duplicate responses were removed.
    	    
    	    \subsubsection{Informal Caregivers} 
    	    
        	    Ninety two percent of the respondents are female, younger than 60 y-o (74\%), completed higher education (70\%), have mild to moderate visual impairments (54\%), are currently unemployed (43\%), and are the children of the \ac{PwD} (60\%).
            	Sixty-one percent live in the same house as the \ac{PwD}.
            	The assessment of perceived daily distress, daily distress caused by the \ac{PwD}, and caregivers' work with the \ac{PwD} reveals that they find themselves in distress and overburdened with the work (median of 4 in 5 for each).
            		
            	Caregivers' stress caused by \ac{PwD} is statistically significantly different across different employment statuses (${\chi}^2(3)\!=\!20.508, \textit{p}\!<\!.00001$). 
            	Post-hoc analysis reveals a statistically significant difference between caregivers with full-time jobs and those that are full-time caregivers ($\textit{p}\!=\!.002$).
            	Daily distress perceived by the caregivers is statistically significantly different across different stages of the disease (${\chi}^2(2)\!=\!13.452, p\!=\!.001$). 
            	Post-hoc analysis reveals a statistically significant difference between mild and severe stages ($\textit{p}=.002$), with caregivers of \ac{PwD} in a severe stage being more likely than those at mild stages to experience daily stress.
            	
            	The amount of work with the \ac{PwD} is also statistically significantly different across the different stages of the disease (${\chi}^2(2)\!=\!40.433, \textit{p}\!<\!.00001$). 
            	Post-hoc analysis reveals a statistically significant difference for all the groups: mild \textit{vs} moderate ($\textit{p}\!=\!.003$), mild \textit{vs} severe ($\textit{p}\!<\!.00001$), and moderate \textit{vs} severe ($\textit{p}\!<\!.00001$).
            	Perceived daily distress, stress caused by the \ac{PwD}, and caregivers' amount of work in the group of caregivers living with the \ac{PwD} are statistically significantly higher than for the group that does not live with the \ac{PwD} ($U\!=\!47459.5, p\!=\!.038$; $U\!=\!53511, \textit{p}\!<\!.00001$; $U\!=\!52393, \textit{p}\!<\!.00001$).
        		
        	    The majority of caregivers are familiar with technology: they use smartphones daily, and computers/tablet/iPad five to six days a week.
        	   
            \subsubsection{\ac{PwD}} 
            
                Sixty seven percent are female, older than 70 y-o (89\%), and have completed at least high school education (67\%).
            	They have mainly mild to moderate visual impairments, such as cataracts, hyperopia, and myopia (61\%).
            	They attend multiple medical specialties, such as neurology (51\%), psychiatry (20\%), psychology (and neuropsychology) (13\%), among others.
            	Caregivers indicate that the \ac{PwD} have a clinical diagnosis of Alzheimer's (47\%), diagnosed for over three years (54\%), and are at a perceived moderate stage of the disease (48\%).
            	    
            	They live mainly at a family home (85\%) with at least one person living with the \ac{PwD} (83\%).
            	The assessment of the level of independence of the \ac{PwD} for different types of activities reveals that they: a) do not cook independently; b) are able to perform activities of daily living but with some help (e.g., dressing); c) are able to move from short to long distances only with the help of third parties; and d) perform some entertainment activities only with the help of third parties (e.g., watching television).
            
        \subsection{Reminiscence Therapy}
        
            Sixty-five percent of the \ac{PwD} are performing reminiscence therapy, while 23\% stopped doing it due to increased detachment of the \ac{PwD} (66\%), lack of perceived usefulness or \ac{PwD} inability to attend sessions (25\%), lack of material (10\%), and lack of time (7\%). 
    		The remaining ones (12\%) never performed reminiscence therapy, as it was prevented by detachment of the \ac{PwD} (66\%), lack of perceived usefulness or \ac{PwD} not being able to participate (25\%), lack of time (8\%), and lack of materials (7\%).

    		We analyzed whether performing (or not) reminiscence therapy was associated with the overall daily stress, daily stress and daily work due to the \ac{PwD}, employment, age, and education of caregivers, as well as the age, education and stage of the disease of the \ac{PwD}.
    		There were no significant relationship for all variables except for caregiver's daily work, \ac{PwD}'s level of education, and disease's stage.
    		
    		The daily amount of work of the caregiver due to the \ac{PwD} is statistically significantly different across those that perform (or not) the reminiscence therapy ($\chi^2(8)\!=\!19.694, \textit{p}\!=\!0.009$). 
    		Post-hoc analysis revealed, as expected, that when the amount of daily work with the patient is higher, therapy is performed less often.
    		Performing reminiscence therapy with a moderate amount of daily work with the \ac{PwD} is significantly more than expected, while those performing therapy with a higher daily work is significantly less than expected.
    		For those that reported a low or moderate amount of work (levels 1 to 3), 67\% currently apply reminiscence therapy, 19\% already did but stopped, and 14\% never did.
    		When considering a higher amount of work (levels 4 and 5), the ones performing therapy reduced to 57\%, while the ones that stopped performing therapy increased to 29\%.
    	
    		There were a statistically significant relation between performing therapy and the stage of the disease of the \ac{PwD} ($\chi^2(6)\!=\!54.461, \textit{p}\!<\!.00001$).
    		Post-hoc analysis suggests that \ac{PwD} in initial stages of the disease perform therapy often: mild (81\%), moderate (74.9\%), and severe (47.1\%).
    		As the disease progresses, there is an increase in the number of \ac{PwD} who stop performing therapy: mild (7.1\%), moderate (16.3\%), and severe (37.1\%).
    		
    		Those performing reminiscence therapy with a moderate stage of dementia are statistically significantly more than expected, while those in a severe stage are statistically significantly less than expected.
    		For those that already performed therapy and the \ac{PwD} has a moderate stage of dementia are statistically significantly less than expected, while those in a severe stage are statistically significantly more.
    	    There were also a statistically significant relation between performing therapy and the education level of the \ac{PwD} ($\chi^2(6)\!=\!20.329, \textit{p}\!=\!0.002$).
    		Post-hoc analysis suggests that those that never performed reminiscence therapy and have a high school degree are significantly more than expected, while the ones with a middle school degree are significantly less.
        
        \subsubsection{Session’s Characterization}
        
            For our analysis, we considered answers from caregivers that were performing therapy and those who already performed.
    		Sessions mainly occur at the home of \ac{PwD} or relatives (92\%) with the attendance of the caregiver (62\%) and relatives (62\%) besides the \ac{PwD}, a sessions' frequency per month of either more than 10 times (44\%) or 1-2 times (23\%), and an average session duration of less than 30 minutes (66\%).
    		When reviewing the past, one can expect that some communication between the \ac{PwD} and remaining participants takes place.
    		In fact, it occurs: in some cases (38\%), a lot (32\%), and little (23\%).
    
    		In the course of each session, different multimedia could be used to elicit memories: image, music, video, or combinations among those.
    		Caregivers use as stimuli mainly images (55\%) or music (37\%), and only 18\% uses videos; they are usually personal, i.e., multimedia that is related to the \ac{PwD} and their life: images (88\%), videos (73\%), and songs (63\%).
    		
    		Images are provided by relatives (81\%), the \ac{PwD} itself (44\%), and friends (12\%). 
    		They are mainly available as printed pictures in physical albums (96\%), but they are also stored in the smartphone/tablet gallery (41\%), or in the computer (27\%).
    		Music is provided by relatives (66\%), Internet through music streaming services (46\%), and the \ac{PwD} (32\%).
    		Songs are usually stored in tapes/CD (57\%), streaming services on the Internet (48\%), and the computer, smartphone or tablet (45\%).
    		Videos are provided by relatives (83\%), \ac{PwD} (24\%), and friends (21\%).
    		They are stored in the computer, smartphone or tablet (71\%), and tapes/DVD (43\%). 
    		
        \subsubsection{\ac{PwD} Emotional Reactions}
        	
            When reviewing our memories, it is common to have different emotional reactions - either positive or negative - depending on the memory in question.
			Regarding the emotional reactions that occur during sessions, caregivers reported that they either perceive what the \ac{PwD} is feeling (62\%) or the \ac{PwD} is able to describe what he/she is feeling (46\%).
			Considering the different multimedia used within sessions, images (66\%) and music (55\%) lead to some kind of positive reaction (e.g., enthusiasm, joy, etc.) from \ac{PwD}.
			Negative emotional reactions occurs during sessions (64\%).
			
			The main situations in which \acp{PwD} experience negative emotions are: i) related to cognitive-impairments (58\%) - not being able to remember something (e.g., when the memories do not line up with the reality), confusion (e.g., \ac{PwD} is able to recognize a person in the image but not being able to recall his/her name), and the inability to express what he/she is feeling; ii) autobiographical memories (44\%) - remembering (or seeing pictures of) loved ones that have passed away, and bad memories from their lives (e.g., abusive parents or spouse, or a divorce).
			While experiencing negative emotions, \ac{PwD}'s have multiple reactions: behavioral, i.e., getting agitated, aggressive, detached, and anxious (56\%), verbal (49\%), and emotional, mainly through crying (47\%).
			Following, we present some quotes of the informal caregivers that exemplify each of the situations reported.
        
                    \begin{table}[!t]
	        \small
	        \renewcommand{\arraystretch}{1.2}
	        \captionsetup{labelformat=empty} 
    	    \begin{tabular}{p{160mm}}
    	        \textbf{Not being able to remember something:}\\
    	        \toprule
    	        \textit{``When memories don't line up with reality.''} [id-55]\\
			    \textit{``Frustrated and angry if he can't remember.''} [id-77]\\
			    \textit{``When she thinks someone who has died is in the house, she gets frustrated and I can see she's scared..''} [id-107]\\
			    \textit{``Does not recognize that both parents passed away about 20 years ago, and cries because she is worried about parents being alone without her not visiting them.''} [id-166]\\
			    \textit{``He began to react negatively to photos of him and his siblings as children, having a fixed delusion that they were starving.''} [id-347]\\
			    \textit{``Gets confused between actual and imaginary. The further back in time, the better.''} [id-394]\\
			    \textit{``When he doesn't remember what happened, he becomes aggressive with the caregivers.''} [id-601]\\ 
			    \\
			    \textbf{Confusion:}\\
			    \toprule
			    \textit{``If confused, he's upset that he's confused.''} [id-99]\\
			    \textit{``When she gets confused with names relating to a photo, but can remember why it was taken.''} [id-164]\\
    	        \textit{``After viewing photos she looks for the person in her home then gets upset they are not there.''} [id-210]\\
    	        \textit{``Gets very upset because he can't remember his grandchilds name. OR can't believe all of his 8 siblings are passed and he is the last one left. Gets angry at himself bc he knows he should know. :( :(''} [id-245]\\
			    \textit{``A mirror outside her bedroom door had to be removed, it scared her and made her angry.''} [id-270]\\ 
			    \\
			    \textbf{Inability to express what he/she is feeling:}\\
			    \toprule
    	        \textit{``When she can't find words to describe her memories, she gets very frustrated, agitated, angry.''} [id-33]\\
			    \textit{``Cannot express what she feels/thinks.''} [id-47]\\
			    \textit{``Frustrated with being unable to communicate and understand.''} [id-312]\\
			    \\
			    \textbf{Remembering loved ones that passed away:}\\
                \toprule    	      
			    \textit{``Gets sad when seeing photos of his deceased wife and sister.''} [id-23]\\
			    \textit{``Family members that have passed away.''} [id-72]\\
			    \textit{``Cries over family gone.''} [id-142]\\
			    \textit{``Death of loved ones and animals.''} [id-263]\\
			    \textit{``Loved ones that have passed away (husband, mother, father, brother, friends).''} [id-274]\\
			    \\
                \textbf{Bad memories from their life:}\\
                \toprule
                \textit{``Memories of parents' drinking and gambling, memories of their divorce and son's drug use.''} [id-253]\\
			    \textit{``Angry and upset and very agitated that people in her past treated her badly.''} [id-264]\\
			    \textit{``Mama has many good memories but it is still affected by memories of her parents, who were abusive to her. She has nightmares sometimes that her alcoholic father may return to bother her.''} [id-280]\\
			    \textit{``Is reminded of an unpleasant memory.''} [id-336]\\
			    \textit{``Remembering father hitting his mother.''} [id-369] 
		\end{tabular}
        \end{table}
            
            In order to cope with the negative emotional reactions, caregivers adopt different strategies:
		    verbal actions toward the \ac{PwD} (67\%), change environment/theme/activity (50.00\%), and non-verbal actions toward the \ac{PwD} (21\%).
		    We can see in the following quotes that caregivers usually talk with the \ac{PwD}, trying to calm them down by rationalizing and validating what the \ac{PwD} are feeling, as well as helping them deal with the triggered memories. 
		    However, sometimes, they opt to avoid continuing to talk about specific themes to not worsen the \ac{PwD}'s (or they own) emotional state.
                    \begin{table}[!h]
	        \small
	        \renewcommand{\arraystretch}{1.2}
	        \captionsetup{labelformat=empty} 
    	    \begin{tabular}{p{160mm}}
    	        \textbf{Verbal actions toward the \ac{PwD}:}\\
    	        \toprule
    	        \textit{``Cuddle, talk, change conversation.''} [id-26]\\
			    \textit{``Try to stay calm, explain gently what she has forgotten or just agree if it doesn't matter.''} [id-111]\\
			    \textit{``Validate the feeling, then try to distract.''} [id-168]\\
			    \textit{``Talk. Sometimes I get upset too :(''} [id-178]\\
			    \textit{``I gently coax them and lead them to come up with the answer another way.''} [id-221]\\
			    \textit{``I Reassure him, change the subject and talk of other memories or other people..''} [id-246]\\ 
			    \textit{``Help her by telling her a memorable moment of that person to help her remember.''} [id-251]\\
			    \textit{``Try prompting, if that doesn't work then distracting/changing.''} [id-257]\\
			    \\
			    \textbf{Change the environment/theme/activity:}\\
			    \toprule
			    \textit{``I stop showing photos and change the subject to something else, most of the time it helps to calm her, but if it continue I walk away out of the room and come back minutes later, then she is good and friendly again.''} [id-6]\\
	            \textit{``Continue looking at other pictures.''} [id-43]\\
			    \textit{``Refocus on positive memory from childhood, family events or activities, person used to enjoy.''} [id-161]\\
			    \textit{``It makes me feel sad. I usually put the photos away and we try again later, in conversation, sometimes with and without photos.''} [id-187]\\
			    \textit{``Giving comfort and appreciation also distracting from topic.''} [id-235]\\
			    \textit{``Change subject, find another photo.''} [id-301]\\
			    \textit{``Distraction such as going for short walk or showing different photo.''} [id-318]\\
			    \\
			    \textbf{Non-verbal actions toward the \ac{PwD}}\\
			    \toprule
			    \textit{``Try to change subject, put my arm around him, hug him.''} [id-7]\\
			    \textit{``Supportive, hugging.''} [id-132]\\
			    \textit{``Just smile and change the subject.''} [id-249]\\
			    \textit{``Console them, hug them.''} [id-323]\\
			    \textit{``Leave her alone.''} [id-339]\\
			    \textit{``I let her talk. Tell her she is allowed to feel this way. Hug her or hold hands.''} [id-396]
	        \end{tabular}
        \end{table}
        
        \subsubsection{Life Story Book}
        
            \acfp{LSB} are frequently used as a reminiscence tool to support recollecting autobiographical memories \cite{Elfrink2018}.
		    They are the result of a life review process performed by the \ac{PwD} with the support of staff and family members to build a personal biography that acts as a bridge to the past and a connection to the present.
		    It can be a very time-consuming task since it may require help from third parties to collect the information, which may not always be feasible or practical.
		    
		    Only 23\% of the caregivers have currently a \ac{LSB} of the \ac{PwD}, but the majority of the remaining ones are interested in having one (80\%).
		    Although we have not collected data to directly answer why most caregivers do not have an \ac{LSB} of the \ac{PwD}, as we can see in the following quotes, our open-end questions allow us to hypothesise that it is due to the amount of work (and stress) caregivers already have in their daily live.
		    
		            \begin{table}[!h]
	        \small	        
	        \renewcommand{\arraystretch}{1.2}
	        \captionsetup{labelformat=empty} 
    	    \begin{tabular}{p{160mm}}
    	        \textbf{Why caregivers do not create an \ac{LSB}:}\\
    	        \toprule
    	        \textit{``Caregivers need more support. These ideas are helpful to the person with dementia, and indirectly to caregivers who find joy when their loved one is happy. Yet, way too much focus is on helping the person with dementia. As a caregiver I am very lonely and exhausted.''} [id-78]\\
    	        \textit{``Some easy way to sort, keep, file items would be appreciated. Since I am the caregiver 24 hours, take care of the house and the yard, and have to find time for myself to stay healthy, the Memory Book is a hit and miss.''} [id-95]\\
    	        \textit{``Many pictures react with her differently but it is hard to have time to organise memory books or to keep her interest in it. I tried to get family involved but they actually got bored of it quicker than mum!!! So, in the end, she gave up too. I want to carry on and do it for her to see later but I just don't have the time working and then going home.''} [id-264]\\
    	        \textit{``We are so busy just being able to get food made and to medical appointments that takes up all the energy for the day. So far, no time for hobbies, just TV with no effort involved.''} [id-353]
        	\end{tabular}
            \end{table}
		    
		    Considering those that already have the \ac{LSB} or are interested in having one, they mainly prefer to store it the old fashioned way, i.e., physically, such as an album of photos (82\%).
		    Eighty-two percent of caregivers had (or want) to have help in the process of creating the \ac{LSB}, and usually they spend (or are willing to spent) up to 10 hours in this process (60\%).
		    The information is obtained in-person (81\%), by e-mail (24\%), and through social networks (23\%).
		    Finally, caregivers indicate that relatives (91\%), \ac{PwD} (62\%), friends (57\%), and the caregivers themselves (49\%) should have access to the \ac{LSB}.
        
        \subsection{Interest in Technological Functionalities}

            Caregivers are very interested in all the technological functionalities (median of 4 in 5 for each) that may help to revisit older memories from their loved ones or to create an \ac{LSB}, namely: i) automatically retrieving new images; ii) retrieving life-based images (of the \ac{PwD}), iii) retrieving interest-based images (e.g., favorite artist, hobbies); iv) adapting current session (based on a negative emotional reaction to an image); v) adapting future sessions (based on the negative emotional reactions that occurred).
        	    
            We further analyzed if the interest in any of the functionalities is associated with currently performing (or  not) the reminiscence therapy, already having a \ac{LSB} or interest in having one, and finally, the overall daily stress, daily stress and daily work due to the \ac{PwD}.
    
                \subsubsection{Automatically retrieving new images}
                    The interest in creating \ac{LSB} has a significant effect on the level of interest in this functionality (${\chi}^2(2)\!=81.234\!, \textit{p}\!<.00001\!$).
                    Post-hoc analysis reveals a statistically significant difference for all pairs: do not want \textit{vs} want ($\textit{p}\!<.00001\!$), do not want \textit{vs} maybe want ($\textit{p}\!<.00001\!$), and want \textit{vs} maybe want ($\textit{p}\!<.00001\!$). 
                
                \subsubsection{Retrieving life-based images}
                    The interest in creating a \ac{LSB} has a significant effect on the interest in this functionality (${\chi}^2(2)\!=74.407\!, \textit{p}\!<.00001\!$).
                    Post-hoc analysis reveals a statistically significant difference for all pairs: do not want \textit{vs} want ($\textit{p}\!<.00001\!$), do not want \textit{vs} maybe want ($\textit{p}\!<.00001\!$), and want \textit{vs} maybe want ($\textit{p}\!=.001\!$).
                    Spearman's rank-order correlation shows a positive correlation between the interest in the functionality and the overall daily stress (${r}_s(601)\!=.100, \textit{p}\!=.014\!$), and the amount of work with the \ac{PwD} (${r}_s(601)\!=.108, \textit{p}\!=.0.008\!$).
                
                \subsubsection{Retrieving interest-based images} 
                    The interest in creating a \ac{LSB} has a significant effect on the level of interest in this functionality (${\chi}^2(2)\!=77.558\!, \textit{p}\!<.00001\!$).
                    Post-hoc analysis reveals a statistically significant difference for all pairs: do not want \textit{vs} want ($\textit{p}\!<.00001\!$), do not want \textit{vs} maybe want ($\textit{p}\!<.00001\!$), and want \textit{vs} maybe want ($\textit{p}\!=.002\!$).
                    A Spearman's rank-order correlation determined that there is a positive correlation between the amount of work of the caregiver with the \ac{PwD} (${r}_s(601)\!=.083, \textit{p}\!=.040\!$) and level of interest in the functionality.
                
                \subsubsection{Adapting current session} 
                    The interest in creating a \ac{LSB} has a significant effect on the level of interest in this functionality (${\chi}^2(2)\!=50.294\!, \textit{p}\!<.00001\!$).
                    Post-hoc analysis reveals a statistically significant difference for two of the pairs: do not want \textit{vs} want (${\chi}^2(2)\!=-98.201, \textit{p}\!<.00001\!$), and do not want \textit{vs} maybe want (${\chi}^2(2)\!=-133.992, \textit{p}\!<.00001\!$).
                    The Spearman's rank-order correlation shows a positive correlation between the amount of work of caregivers with the \ac{PwD} (${r}_s(601)\!=.099, \textit{p}\!=.015\!$) and level of interest in the functionality.
                
                \subsubsection{Adapting future sessions}
                    The interest in creating a \ac{LSB} has a significant effect on the level of interest in this functionality (${\chi}^2(2)\!=38.561\!, \textit{p}\!<.00001\!$).
                    A pairwise post-hoc Dunn test with Bonferroni adjustments was significant for the pairs: do not want \textit{vs} want ($\textit{p}\!<.00001\!$), and do not want \textit{vs} maybe want ($\textit{p}\!<.00001\!$).
     
        \null
        Conducting reminiscence therapy sessions does not seem to be associated with the level of interest in any of the technological functionalities presented.
        Caregivers that are not interested in creating an \ac{LSB} are less interested in the aforementioned technological functionalities than those interested in having one, and in the majority, those that want to create an \ac{LSB} are slightly more interested than those that \textit{might} want to create the \ac{LSB}.
        With a higher overall stress of the caregiver, the interest in \textit{Retrieving life-based images} appears to increase.
        A higher amount of work of the caregiver with the \ac{PwD} increases the interest in \textit{Retrieving life-based images}, \textit{Retrieving interest-based images}, and \textit{Adapting current session}.
\section{Survey 2}
    
    In this section, we present our results regarding sample characteristics (of caregivers), how reminiscence sessions are conducted, and how the results of reminiscence sessions (and overall health of the \ac{PwD}) are evaluated in follow-up medical appointments.
    Finally, we assess the level of interest in technological functionalities to help in the care of \ac{PwD} (management of patients, information related to sessions carried out, management of the multimedia material to be used in sessions, and which help tools should be available), and how this interest differs between formal and informal caregivers.
    
    The privacy of the information exchanged is very important, especially when other people are involved.
    Beyond ethical issues, the non-guarantee of privacy may even hinder the embrace of new technology from caregivers.
    As such, we also address how comfortable caregivers would be inserting personal content on technological solutions (assuming that only authorized people would be able to consult the data).


    \subsection{Informal Caregivers}
    
        Following, we detail the results obtained for the informal caregivers.
        Where relevant, we compare the results with those of the previous study.
    
        \subsubsection{Sample Characteristics}
    
            A total of 102 informal caregivers worldwide answered our survey.
            Ninety two percent were female, younger than 70 y-o (90\%), completed higher education (65\%), and have mild to moderate visual impairments (58\%).
            They usually take care of one \ac{PwD} (87\%) that is either they spouse (39\%) or a relative (53\%). 
            Usually, the \ac{PwD} they look after are in a state that is perceived as severe (61\%).
            Finally, the average number of hours in the day that the caregiver spend caring for the \ac{PwD} is above 8 (equivalent to a full-time job)  (60\%).
                
            Most caregivers are familiar with technology: they use mobile phones five to six days a week, and computers/tablet/iPad three to four days a week.
            
        \subsubsection{Reminiscence Therapy}
        
            Forty-one percent of caregivers perform reminiscence therapy with the \ac{PwD} they care for.
            The number of sessions conducted per month is either more than 10 (43\%), 3-4 (21\%) or 1-2 (21\%).
            Sessions occur mainly at home (88\%), with an average duration of less than 30 minutes (74\%).
            
            It is worth noting that the performance of the reminiscence sessions decreased compared to the previous survey (about 20\%).
            We believe it is due to the fact that informal caregivers are increasingly overburdened and subject to high levels of stress, which may have been exacerbated by the pandemic situation experienced at the time of data collection.
            Moreover, we believe the existing confinements due to the pandemic increased the isolation of the \ac{PwD}, decreasing the contact between them and their relatives, both at home and in the institutions, which conditioned the realization of the therapy, and made it difficult to gather materials for the sessions.
            
            Images (88\%), music (67\%), and videos (43\%) are used as triggers to stimulate the memory of \ac{PwD}, and are mostly personal (95\%).
            They are related to the relatives of the \ac{PwD} (93\%), jobs and places the person visited (62\% each), places where the person lived (57\%), pets (52\%), vehicles they had (33\%), among others (23\%), such as parties/happy moments, favorite activities, childhood memories, food, and dance.
            Such materials are provided mainly by relatives of the \ac{PwD} (81\%), friends (29\%), the person itself (21\%), or  gathered from the internet (40\%).
            Caregivers would like to use other multimedia more often: videos (52\%), images (38\%), and music/text (29\% each).
            Images and music usually elicit positive emotional reactions in the \ac{PwD} (71\% each), while images (26\%) and videos/text (24\% each) usually elicit negative emotional reactions.
        
            Caregivers update the material on a daily (29\%),  monthly (26\%), or weekly (17\%) basis.
            The main reasons that keep them from updating the material more often are: the detachment of the \ac{PwD} (50\%), lack of time (30\%), and lack of materials (20\%).
            Seventeen percent of the caregivers never update the material.
            We also believe that the caregivers' overload and stress, as well as the pandemic situation has affected not only the completion of the reminiscence therapy, but its quality when performed since caregivers did not have the time or materials available to diversify the following sessions. 
            
            Regarding the information collected by the caregivers from each session, they register the materials that elicited positive emotional reactions (57\%), materials that helped to recall past memories (55\%), the emotional state of the \ac{PwD} both before and after the session (38\% each), and the material that elicited negative emotional reactions (19\%).
            There are no other types of information that caregivers would like to include in their notes about each session.
            Fourteen percent of caregivers do not collect any information.
            
        \subsubsection{Follow-up Medical Appointments}
    
            
            Thirty-seven percent of surveyed caregivers take the \ac{PwD} to follow-up medical appointment, in particular they accompany the \ac{PwD} on a daily (32\%) or weekly (16\%) basis.
            Thirty-four percent of caregivers reported that the \ac{PwD} do not go to follow-up appointments despite having indicated they did in the previous question. 
            We believe this is due to not having immediately understood the concept of follow-up appointment.
            Caregivers report they do not attend the appointments more often due to lack of time.
        
            The emotional reactions and emotional well-being of the \ac{PwD} is usually analysed during the appointments (64\% each), as well as the results from previous reminiscence sessions (48\%).
            Additionally, caregivers perform other follow-up of \ac{PwD} besides medical consultations or reminiscence therapy, such as cognitive stimulation, walks in places \ac{PwD} used to enjoy, art therapy, shopping, visiting places where \ac{PwD} lived, psychomotricity activities, daily contact with the \ac{PwD}, and gathering relevant information through collaborators/health team.
            
        \subsubsection{Interest in Technological Functionalities}
        
            Thirty three percent of caregivers take advantage of technology to take care of \ac{PwD}, such as audio books, blue laser light, camera, computer, Internet, phone, radio, tablet/iPad, television, and videos.
                
            Considering the different technological functionalities for the management and adaptation of reminiscence sessions, informal caregivers have a moderate interest for the most part (median of 3 in 5 for each).
            This apparent decrease in interest (compared to our previous survey) will be addressed in the follow-up interviews.
            The actual performance of the reminiscence therapy has not proved to be statistically significant for any of the functionalities in question.
                
            Caregivers indicate that they are comfortable introducing personal content into a technological system, as long as the system guarantees maximum privacy and security.
            
     \subsection{Formal Caregivers}
    
        Following, we present our results regarding formal caregivers.
        
        \subsubsection{Sample Characteristics}
        
            A total of 34 formal caregivers answered our survey. 
            They were mostly female (85\%), younger than 50 y-o (64\%), completed higher education (85\%), and do not have any visual impairments (65\%).
            They take care of more than seven \ac{PwD} (76\%), and at least one of the \ac{PwD} is in a severe state of dementia (74\%).
            
            The majority of caregivers are familiar with technology: they use mobile phones daily, and computers/tablet/iPad six days a week.
                
        \subsubsection{Reminiscence Therapy}
        
            Formal caregivers perform reminiscence therapy sessions more often than the informal ones (68\% vs 41\%).
            They usually conduct the sessions 3-4 (39\%) or 5-10 (30\%) times per month.
            Sessions occur mainly at the institution (74\%) but in the house of the \ac{PwD} as well (35\%), with an average duration of 30-60 minutes (65\%) or less than 30 minutes (30\%).
            
            Images (95\%), music (87\%), videos (57\%), and text (48\%) are used to stimulate memories, and are both personal (87\%) and general (83\%).
            Caregivers would like to use video (57\%), text (48\%), images (43\%), music (39\%), and other materials (such as objects, arts' material, games, daily activities, and visits to places of interest) more often.
            
            The materials used are related to places where the person lived (96\%), \ac{PwD}'s relatives (87\%), places \ac{PwD} visited (83\%), jobs of the \ac{PwD} (78\%), pets (70\%), and vehicles they had (43\%).
            Caregivers also report that they like to use material related to other topics such as habits, hobbies, favorite food, musicians, television shows, movies and actors, important dates, past and present objects, sports, or plants.
            Materials are obtained through the Internet (87\%), or provided by \ac{PwD}'s relatives (74\%), the \ac{PwD} (57\%), or friends (43\%).
            The material is updated in a monthly (26\%), weekly (22\%), daily or fortnightly (17\% each) basis.
            They do not update it more often due to lack of time (100\%), lack of perceived usefulness for the \ac{PwD} (50\%), and lack of materials or need to repeat the same activities (25\% each).
                
            Caregivers collect information about the emotional state of the \ac{PwD} after sessions (87\%), material that helped to recall past memories or elicited positive emotional reactions (83\% each), material that elicited negative emotional reactions (70\%), \ac{PwD}'s emotional state before sessions (65\%), and duration of each session (48\%).
            Moreover, they also collect the degree of involvement and interest of \ac{PwD}, level of communication, dynamics with caregiver or the group, \ac{PwD} answers during the session, as well as periodic assessments of cognitive functions.
            
            Finally, they show interest in being able to collect the intensity level of emotional reactions (positive and negative) to specific stimuli, record permission for assessments, how the \ac{PwD} was during the session (refusals, non-verbal language, behavior, relationship with therapist, initiative in contact), and in-session responses that might be relevant in future sessions.
        
        \subsubsection{Follow-up Medical Appointments}
        
            Fifty-three percent of the formal caregivers conduct follow-up medical appointments with the \ac{PwD} on a daily (28\%) or weekly (28\%) basis.
            Caregivers report they do not conduct appointments more often because of the costs associated.
                
            The emotional well-being of the \ac{PwD} is almost always evaluated at the appointments (93\%), followed by the \ac{PwD}'s emotional reactions (87\%), and the results from previous reminiscence sessions (67\%).
            Other follow-up activities with the \ac{PwD} are conducted (49\%), in particular, cognitive stimulation, movement sessions, sensory stimulation sessions, hygiene/food care, caregiver support with family members and joint sessions (\ac{PwD} and family), psychotherapy, and games.
        
        \vspace*{0.25cm}
        \subsubsection{Interest in Technological Functionalities}
            
            Sixty-five percent of caregivers already use technology to help them in the care of the \ac{PwD}, in particular, the CogWeb application \cite{Cruz2013}, computer, images, music, mobile phones, overhead projector, snoezelen (multi sensory) room, tablet, and videos.
            
            They are interested in all the technological functionalities, in particular, they are very interested in those related to the adaptation and personalization of sessions.
            Their interest is much higher than that of the informal caregivers, which we believe is due to the fact that most of the informal caregivers only take care of one person and do not need to manage several \ac{PwD} at the same time.
            Such difference in the interest across the type of caregivers is statistically significant for all functionalities:
            
            \null
            \noindent\textbf{Management of \ac{PwD}:}
        	    \begin{itemize}\setlength\itemsep{-0.15em}
        	   	    \item \textit{Creating reminders} (${U}\!=\!1063.50\!, \textit{p}\!=\!.001\!$)
        	   	    \item \textit{Notifying other caregivers} (${U}\!=\!1089.00\!, \textit{p}\!=\!.001\!$)
        	   	    \item \textit{Sharing care with others} (${U}\!=\!1085.50\!, \textit{p}\!=\!.001\!$)
                \end{itemize}
        	\noindent\textbf{Reminiscence sessions:}
        	    \begin{itemize}\setlength\itemsep{-0.15em}
        	        \item \textit{Automatically adapting the session content} based on the biographical information of the \ac{PwD} (${U}\!=\!758.50\!, \textit{p}\!<\!.00001\!$)
        	        \item  \textit{Automatically adapting the session content} based on the emotional reactions of the \ac{PwD} to the materials presented (${U}\!=\!862.50\!, \textit{p}\!<\!.00001\!$)
        	        \item \textit{Consulting the materials used} in each therapy session (${U}\!=\!801.00\!, \textit{p}\!<\!.00001\!$)
        	        \item \textit{Consulting the history} of caregiver's sessions (${U}\!=\!940.00\!,  \textit{p}\!<\!.00001\!$)
        	        \item \textit{Consulting the history} of the \ac{PwD}'s sessions even if performed by others (${U}\!=\!1007.50\!, \textit{p}\!<\!.00001\!$)
        	    \end{itemize}
        	   
        	\noindent\textbf{Material used in the therapy sessions:}
        	    \begin{itemize}\setlength\itemsep{-0.15em}
        	        \item \textit{Inserting new material} (${U}\!=\!862.50\!, \textit{p}\!<\!.00001\!$)
        	        \item \textit{Inserting material topics} (${U}\!=\!891.00\!, \textit{p}\!<\!.00001\!$)
        	        \item \textit{Creating a favorites section} in each patient's material (${U}\!=\!937.00\!, \textit{p}\!<\!.00001\!$)
        	    \end{itemize}
            \noindent\textbf{Help tools for the users:}
        	    \begin{itemize}\setlength\itemsep{-0.15em}
        	        \item \textit{Email/Telephone} (${U}\!=\!1276.50\!, \textit{p}\!=\!.018\!$)
        	        \item \textit{Help within application} (${U}\!=\!1168.50\!, \textit{p}\!=\!.004\!$)
        	        \item \textit{Instructions manual} (${U}\!=\!1231.00\!, \textit{p}\!=\!.010\!$)
        	 \end{itemize}
        	        
            Similarly to the informal caregivers, formal caregivers currently performing (or not) the reminiscence therapy has not proved to be statistically significant for any of the functionalities.
            They are also comfortable in introducing personal content of the \ac{PwD} they care for into a technological system, as long as the system guarantees maximum privacy and security.
    \renewenvironment{quotation}
               {\list{}{\listparindent=0pt
                        \itemindent    \listparindent
                        \leftmargin=14pt
                        \rightmargin=14pt
                        \topsep=8pt
                    }%
                \item\relax}
               {\endlist}

    \section{Follow-up Interviews and Focus Groups}

        In this section, we present the main findings gathered from the interviews and focus groups we conducted.
        Although these were analyzed separately, the results are focused on the same aspects, thus we opted to present them together.
        
        Our aim was to attain more in-depth information regarding the following topics: i) identify which materials caregivers used, how they acquire such materials, and the role of personalization in reminiscence sessions; ii) identify mechanisms to help in the creation of sessions; iii) understand which information is useful to be registered from the sessions performed; finally, iv) how to support the communication and share of \ac{PwD} care among caregivers.
        
        \subsection{Sample Characteristics}
        
        A total of 2 informal and 18 formal caregivers participated in the interviews and focus groups.
        
        \subsubsection{Semi-structured Interviews}
            
            We conducted two rounds of interviews.
            Two informal female caregivers (Portuguese (IN-1) and Brazilian (IN-2)), and a formal caregiver (Portuguese psychologist (F-1)) participated in the first round.
            Caregiver IN-1 cares for her spouse, and IN-2 for his mother.
            Six Portuguese formal female caregivers participated in the second round: five psychologists (F-2, F-3, F-4, F-6, F-7), and a neuropsychologist (F-5).
            They work in dementia's field at Alzheimer Portugal (F-3, F-4, F-6), Centro de Apoio Alzheimer Viseu (F-1, F-2), Hospital Santa Maria (F-5), and Santa Casa da Misericórdia de Lisboa (F-7).
            
            Note that caregivers F-1 and F-2 are the same person, but given that the interviews were conducted at different periods of time and contemplated different aspects, we chose to refer to her as if it were two caregivers.
            
        \subsubsection{Focus Groups}
            
            Two focus groups were carried out: one in Irmãs hospitaleiras - Casa de Saúde de Idanha and the other in the Santa Casa da Misericórdia de Lisboa.
            Eight formal caregivers involved in the care of \ac{PwD} participated in the first focus group (F-8, F-9, F-10, F-11, F-12, F-13).
            Five formal caregivers participated in the second focus group (F-14, F-15, F-16, F-17, F-18).
            They were neurologists, neuropsychologists, nurses, occupational therapists, physical therapists, and social workers.
            They were all Portuguese.
        
        \subsubsection{Multimedia Material}
        
            The use of multimedia material is very important to the reminiscence therapy. 
            Caregivers normally use photos and music to trigger the \ac{PwD} memories, and whenever possible, ensuring an environment the \ac{PwD} enjoys while helping in the process of reminiscing.
            
            Personalized multimedia of the \ac{PwD} life and things they enjoy are usually preferred by caregivers for that person's reminiscence therapy.
            In general, personalized information is a better memory trigger than generic multimedia because it was part of that person’s life. 
            However, and as pointed out by caregivers, the use of both is useful and desirable:
            \begin{quotation}
                \textit{``I often start by showing generic images and songs and then I'll try to narrow it down a little bit. If I know there's a person who likes some particular ones or it make a greater impact on them, I make it more personal."}~[F-1]
                
                \textit{``Use some general images of animals, children, or families but with a positive connotation. Unless I knew that the person really likes snakes, I would never show a picture with snakes, because the probability of triggering a negative reaction is very high."}~[F-3]
                
                \textit{``I would add the part about interests and motivations besides the family background because I think it complements a lot how we implement the intervention."}~[F-9]
                
            \end{quotation}
            
            Regarding the update of the materials used in therapy, informal caregivers (when possible) resort to other family members to obtain new images (e.g., through messaging platforms).
            When the therapy is performed at an institutional level, family members usually provide materials over time, but sometimes they are not able to and it becomes difﬁcult for the formal caregiver to update and personalize the materials to be used, as it depends a lot on the experience and knowledge of the formal caregiver:
            \begin{quotation}
                \textit{``Often through the phone. We have a family group so they keep sending it to me."} [IN-2]
                
                \textit{``It all depends, once again, on the person and the stage they are in. For people who, for example, have a greater degree of forgetfulness in relation to the present reality, sometimes bringing more up-to-date photos will confuse the person even more, won’t it? In other words, it's really a job that has to be very personalised."} [F-1]
                
                \textit{``Sometimes people bring their own photos, and we work on that part. Sometimes there are other people who never bring (photos), and I end up, for example, bringing some pictures of old town, or especially objects that lead people to talk about them, the tasks they did with them, etc."} [F-1]
            \end{quotation}
            
            A major difficulty when intervening with \ac{PwD} is to gather real images of the topics being discussed:
            \begin{quotation}
                \textit{``One difficulty we have is in collecting material because all that exists is very childish and it ends up being a lot of homework. When I prepare sessions, I search for images that are as realistic as possible because otherwise it doesn't create any motivation in people... They feel much more involved than with images of kittens or puppies."} [F-3]
                
                \textit{``If we are talking about a session of food, it should look real, not be a drawing of a lettuce. This may seem a detail, but it's really important, we value this very much: to remove the burden of infatilization of \ac{PwD}."} [F-4]
            \end{quotation}
            
            Most of the caregivers considered relevant the possibility of having the multimedia material catalogued with categories and descriptions, so that it could be organized, and be used as a facilitator in the management of the sessions:
            \begin{quotation}
                \textit{``In other words, it creates a topic of conversation for that picture. I think it is important because it promotes the connection between one thing (the image) and the other (description/categories)."} [IN-1]
        
                \textit{``I would suggest old photographs to be categorized, for example, by decades, 1930s, 1940s, I don't know, something like that."} [F-3].
                
                \textit{``Description... Ah well this makes sense, this makes a lot of sense. Because if they are very significant images to the self it should have a description."} [F-5]
                
                \textit{``The category also helps with searching in the future, doesn't it?... This helps a lot. As a technician, I also use a lot of images stored on the computer and of course it would be really handy to have them organized."} [F-6]
                
                \textit{``This makes perfect sense to us, the more categories the better because it actually makes intervention much easier afterwards."} [F-9]
                
                \textit{``It should be possible to have information regarding the interests of the \ac{PwD}, possibly previous interests, but also current interests, to have this a little bit more level-oriented. In the sense that it would be more convenient to have the information organized, and it would be easier for the various caregivers to access the information in a more directed way."} [F-12]
            \end{quotation}
            
            The use of descriptions accompanying the images might be even useful in other activities:
            \begin{quotation}
                \textit{``Until recently, I used the images that a family had of their relative, they have lots of pictures of him doing lots of actions. Riding a horse, for example. At Christmas, and since he is a person who already has a lot of communication difficulties, it served to stimulate his reading skills, his attention span, in which of the images is actually riding a horse. Then he is very happy because he recognizes that he is the one riding a horse: `Look, that's me riding a horse, interesting. It's written here, and I'm riding a horse'. Yes, I think that this way it becomes much more organized."} [F-6]
            \end{quotation}

            The possibility of editing the information on the images, or even being able to remove them (from the caregiver's or patient's current collection) was well received by caregivers:
            \begin{quotation}
                \textit{``Editing is always good. Sometimes we may want to modify something"} [IN-2]
                
                \textit{``We may end up realizing that the patient actually loves this image and therefore that information can be stored there as well."} [F-4]
                
                \textit{``Let's imagine, I put up x number of images and for some reason in that sequence there is one of them that is really very difficult to work with either because the patient gets very upset or for whatever reason. Or even because we think it is not a good image because it creates more confusion than it is useful. It should be possible to remove them."} [F-5]
            \end{quotation}
            
            Caregivers also showed great interest in the possibility of multimedia content being provided by third parties (e.g., family and friends), as well as sharing existing content between patients/caregivers (whenever appropriate).
            \begin{quotation}
                \textit{``It makes a lot of sense (to be able to make an image public), of course. There are images that are very specific to \ac{PwD} but can also be used with others."} [F-5] 

                \textit{``Let's imagine, a caregiver feels that a particular image brings benefits to the \ac{PwD} they care for, a mechanism to suggest that image to caregivers of other people would be useful."} [F-6]
            \end{quotation}
            
            Privacy is very important for all people, specially when considering older people.
            It is important to note that the share of multimedia content must respect the privacy of the \ac{PwD} and their caregivers:
            \begin{quotation}
                \textit{``For formal caregivers, it's always important that this sharing of multimedia is protected. It protects the caregiver who is working, doesn't it? It also protects the \ac{PwD}."} [F-6]
            \end{quotation}
            
        \subsection{Creation of Reminiscence Therapy Sessions}
        
            With the information collected, and based on existing literature, we inquired the caregivers about three - complementary - technological approaches to support the creation of personalized sessions:
            
            \begin{itemize}\setlength\itemsep{-0.15em}
                \item \textbf{Automatic}: The caregiver would be responsible for selecting the duration and/or number of images to be presented in the session. The choice of images to be shown would be left to the technological solution (considering the \ac{PwD}'s biographical information, \ac{PwD}'s emotional reactions to the stimuli used, emotions conveyed by the stimuli, and finally, caregiver's feedback on the sessions performed so far);
                \item \textbf{Semi-Automatic}: The caregiver would be responsible for the choice of the duration and/or number of images in the session, as well as the topics to be considered. The choice of the images will be similar to the one of the automatic approach, but considering the topics selected. The caregiver would then have the possibility to review the session created and make changes if they want to.
                \item \textbf{Manual}: All the details of the session will be customized by the caregiver. 
            \end{itemize}
            
            Our hypothesis is that the first approach will be particularly helpful for informal caregivers who are overburdened with the care of the \ac{PwD}, and are physically and emotionally worn out.
            We believe the third approach will be more useful for formal caregivers, who need more control in conducting the sessions (based on the different psychological aspects to be worked on in each one).
            
            Finally, we hypothesize that the second approach might be useful for both informal caregivers (who are more proficient in technology and more willing to personalize sessions), and for formal caregivers (who want to speed up the creation of sessions without needing full control in the process).
            As we can see, the approaches presented were well received:
            \begin{quotation}
                \textit{``Talking a little bit about my practice, it's good to have automatic things, no doubt about that, but if it's possible the automatic part should be personalized to the person in question. Always take into account the life history of the person, that is essential for the success of the activities. It doesn't mean that a novelty doesn't bring benefits, sometimes it does. This aspect (of the automatic session) seems interesting to me. As a matter of fact, we also have the possibility of using Cogweb... it's automatic, isn't it? But we still have some kind of choice, we have lots of games to choose from. And it's interesting that they have the same concern that we have, for example, some users have hearing problems and all that, and there is a part where we can select games that are not so dependent on hearing. That already helps us to leave out a number of activities that come to harm."} [F-6]
                
                \textit{``It makes sense (the idea of the automatic session), it even makes sense that there are already some photos available, but it should be possible for people to associate other meaningful photos to the person. It would be useful if they already had some material that would allow them to quickly do a session with that person, even if it is not highly personalized... the ideal would be to be personalized but personalized sessions are not always done according to the experiences and tastes of each person. It would be important not only for the informal caregivers but also for the technicians."} [F-8]
                
                \textit{``It seems important to me that when the themes to be included are chosen automatically, that those that we already use more in the sessions are taken into consideration, for example, the typical dishes of a region."} [F-8]
                
                \textit{``I think that the automatic session would not be the ideal model, but maybe it is the one that will be used in clinical practice, at least in some contexts. That's because with the overload of sessions that sometimes exist, caregivers doesn't have time to program a session for that moment."} [F-12]
                
                \textit{``I think it would be good in each session to have the opportunity to select which images are going to be used. The fact that you add dogs, it doesn't mean that all images of dogs should appear... that allows the therapist to program the session, and think and establish how they want to conduct it."} [F-12]
            \end{quotation}
            
        \subsection{Feedback to be Collected from Sessions}
        
            Collecting feedback from the sessions performed is very important. 
            It allows caregivers to compare sessions, and customize them over time to deliver a better therapy, tailored to the current condition of each \ac{PwD}.
            The possibility of collecting feedback from each session, and access those records at a later date sparked broad interest among caregivers:
            \begin{quotation}
                \textit{``It's important to record the image and the reaction to the image to understand if there was an evolution or retreat."} [IN-1]
                
                \textit{``This is excellent to see her history, all the (therapy) sessions and the evolution of the disease. I appreciate it a lot."} [IN-2]
                
                \textit{``In my way of programming (sessions), I take the time to prepare them, and one of the things I have is an Excel spreadsheet that I created, in which for each session that I prepare, I write down which areas were stimulated, for example. Then, when I go to create the next session, I already have a map of what has been worked on to guide me in what to do next."} [F-1]
                
                \textit{``The last session performed is important because imagine I have 20 people to work with, then I can see who was the one where intervention was done the most time ago, and I would go immediately to that one. If I could order it, even better."} [F-1]
                
                \textit{``It makes perfect sense to me because this is also a way to evaluate the quality of this intervention and where things are going less well. Whether it's because of the session itself (how it's being structured), and there may be differences between the formal and informal caregiver (if the \ac{PwD} in question is doing sessions with both formal and informal caregivers). For the management of the therapy itself this is useful information. If if it's worth continuing, if it's not worth continuing, if we've actually found out that for the patient it's (the intervention) more disruptive than beneficial."} [F-5]
                
                \textit{``It also makes sense to have a summary that serves as a profile of the work I have been doing with the \ac{PwD} and that guides me later in the intervention: when did I start, when did I have last session, how many sessions have I done, what themes have I approached, what themes have I approached that were more successful, what should I not present again as a stimulus (to the \ac{PwD})."} [F-9]
            \end{quotation}
        
            Several suggestions were made and discussed regarding which information (and how it) should be collected.
            The general consensus is that the process should be as simple as possible. 
            It should be possible to review the multimedia content shown during the session, and for each one, collect the \ac{PwD}'s emotional reactions (negative, neutral, and positive). 
            Further granularity in the information collected is only useful in the case of negative emotional reactions.
            Ways to visualize the \ac{PwD}'s evolution over time are also desired.
            \begin{quotation}
                \textit{``The mood graph could be interesting to give us a little bit of that notion. There's something here that I don't know if it's possible to quantify: imagine that we've presented the same image several times and that, for example, the first time, the person said something, but very little.Then the second time, he spoke more and with more memories, and then a third time, evoked even more memories. I don't know if there is any easy way that we can somehow graphically show this. An interesting option would be at the ﬁnal, at the collection of feedback, for example, to show the images that the person has seen and ask the caregiver to select which ones were the most relevant or which ones triggered the most communication.} [F-1]
                
                \textit{``Collect which kind of emotion it triggered and another feedback that is an open field where the person can write some observation... Another suggestion is to collect the feedback not only from the caregiver's subjective input, but from the person him/herself. In other words, if the person liked or disliked the image, and the person him/herself should verbalize that, rather than a subjective `I think the person liked it' evaluation (of the caregiver), is really asking the person if they liked it or didn't like it. This evaluation can also be important."} [F-3]
                
                \textit{``Using the three levels: green, yellow and red to say whether the session went well is important, and it is also important to leave more notes as to what happened for that session to go wrong, in cases where the session goes less well."} [F-5]
                
                \textit{``Start with something very simple, for example, if the session with this image was positive or negative. If the person wrote positive that's it, but if they wrote negative maybe try to have an open field in which they can identify if there was agitation, if there was anxiety..."} [F-6]
                
                \textit{``For example, the person was very sad when you did a session with the wedding images, for example. It's important that this is described so that others can understand that maybe the sessions shouldn't be focus on that at this time."} [F-8]
                
                \textit{``If therapy is an ongoing process, it makes sense to keep track of the evolution of the effects of those therapy sessions. Let's imagine that I love, Tony Carreira or Lenita Gentil, and that this is associated with a very sad memory and that suddenly the person starts having more stressful situations, more agitation. I think this has to be reported and kept for the next person. Normally our teams work in collaboration, so let's imagine, the psychologist plans the profile of what is going to be done in the therapy, and the socio-cultural animator reinforces it and carries it out, and she notices situations in which it is necessary to reconsider because the \ac{PwD} became too agitated, or because, on the contrary, he became too sad, or had insomnia... "} [F-14]
                
                \textit{``I think it could be interesting (to have feedback from sessions), for example, with a grid of the interest, motivation, and participation of the \ac{PwD}. This would allow another caregiver to understand the degree of \ac{PwD}'s motivation and participation."} [F-16]
                
                \textit{``It is interesting because in a session, (the \ac{PwD}) may be in a certain mood, may or may not have engaged (with the session).."} [F-17]
                
                \textit{``It might also be interesting to make an evaluation of how it went, where a balance would be made of what worked and what didn't."} [F-18]
            \end{quotation}
            
            It was also pointed out that the caregiver should not need to input too much information in the feedback of the sessions, as it can be bothersome on a daily basis.
            Feedback should, ideally, be collected at the end of the session, or after consulting the session summary (even days later):
            \begin{quotation}
                \textit{``We are always following the person very closely, moment to moment, and so we leave a little bit that part (of recording the information) for later. So that the person doesn't feel that `Oh, he seems to care more about recording than being with me and talking to me'. People, when they are with us, need to feel that we are `whole' in what we are doing with them.."} [F-1]
                
                \textit{``So I either record it on a little piece of paper next to me or usually, and because I know the people I'm working with very well, at the end of the session I make my record of the session and how the person performed in that task. Well, as long as not too many days go by."} [F-1]
            \end{quotation}
            
            The usefulness of making observations from what happened in the session was also emphasized:
            \begin{quotation}
                \textit{``One thing is the record (of information) of a session, another thing is... I don't know, I'm remembering the psychotherapy sessions: when I'm doing my reflection on a session, I write something about that session and then I leave my observations about what I think is the way to follow (in the next sessions)."} [F-1]
            \end{quotation}

            As a result of the reminiscence therapy sessions performed with the \ac{PwD}, caregivers suggest it should be possible to structure all the information collected to create a book (similar to a \ac{LSB}) to make available to families as a memoir of the \ac{PwD}.
            \begin{quotation}
                \textit{``We would like to be able to print the work done with the person. It would be important for the caregivers to also do this because in the psycho-educational sessions they also bring us the work they do at home with their patients and this is significant for them to share with us and also to share with the other caregivers in these (psycho-educational) sessions. So here the idea is to sistemize and create a book of the memories/reminiscences of the person."} [F-8]
                
                \textit{``To what extent can we not think afterwards about an end product for the person and their relatives. How are we going to give this back since this is so significant because we are going to write things about their life story, their pictures, everything.."} [F-9]
            \end{quotation}
        
        \subsection{Communication/Care Share among Caregivers}
        
            The need to share the care of a \ac{PwD} between caregivers is a reality for both informal caregivers (to feel more supported, less burdened, and less isolated) and formal caregivers (to optimize the relationship with other caregivers who intervene with the person, and improve the care provided).
            \begin{quotation}
                \textit{``And this sharing doesn't have to be only with the healthcare technician, for example, our mutual daughter is not in the country.  I could share and she could remotely follow the work being done with her father, right?"} [IN-1]
            
                \textit{``(The sharing) would certainly make it easier. That's a wonderful idea."} [IN-2]
                
                \textit{``Imagine that the person is in a day center, for example, and that we conduct sessions at our context (office-based intervention), the family caregiver at home can see some images and sessions that we've done with the \ac{PwD}."} [F-3]
                
                \textit{``It makes sense to have multiple people assigned to that person."} [F-5]
                
                \textit{``For example, here in the association we have consultations but then... the person can go to a day center and so the professionals there can also be on board, and it's possible to share information between everyone."} [F-6]
                
                \textit{``For example, it is relevant to know the technician who did some intervention if I can see the intervention. As another technician or as an informal caregiver, right? Because they, for example, can get ideas from the sessions that the technicians did..."} [F-8]
                
                \textit{``It would be important for the informal caregiver to define his/her support network. It can be a professional that is going to do the reminiscence therapy, but there can be other professionals that are already involved and can be important later on for the intervention itself or can provide important points for the intervention itself. For example, a doctor that accompanies the person in a specialty consultation, if there is an institution or a day care center that the person attends...} [F-9]
                
                \textit{``That (the care sharing) could then allow us to do a monitoring of the \ac{PwD} status here after discharge. So we can monitor care because sometimes there are people who we don't lose contact with and the progress of, but there are others that after discharge we don't know about the continuity of care..."} [F-11]
             \end{quotation}  
            
            Caregivers also need to choose which information they want to share and with whom:
              \begin{quotation}
              
              \textit{``Not all caregivers need to have access to everything, because it also depends on their role. One thing is the psychologist who has to do the sessions, the primary caregiver, or the social worker who manages the whole process of that \ac{PwD} and the interconnection with the family, and another thing is the person who just gives him the bath. Then, there is the issue of professional confidentiality, we deal with some very sensitive information of the \ac{PwD}."} [F-14]

              \textit{``I think it would be important that some of the information would not be available to everyone. I'm thinking, for example, that as a result of the application of some psychology tests, or other more specific areas, it wouldn't make sense for it to be available to, for example, geriatric assistants.  There is information that doesn't need to be available to everyone."} [F-18]
            \end{quotation}
            
            Formal caregivers also suggest it should be possible to manage who are the primary caregiver, and be able to transfer the care at any moment:
            \begin{quotation}
                \textit{``That was my question (if you could transfer the primary care to another caregiver), because imagine we have an inpatient here, we start the therapy but then we want the family member (when the \ac{PwD} is discharged) to continue the care."} [F-8]
            \end{quotation}
       
            The use of messages (in an integrated way) in a technological solution attracted the interest for both informal and formal caregivers, but was most welcomed by the formal ones.
            The biggest requirement regarding the use of messaging, particularly between the two types of caregivers but also among formal caregivers, is to ensure that there is the possibility of private messaging:
            \begin{quotation}
                \textit{``Sometimes it might be important to have an exchange of ideas just between two professionals about the person that would not involve others. And if that was kept a record for that person, it means that it would always be accessible there for later on, not forcing caregivers to have to be on their phone or Facebook to share that information just between themselves."} [F-2]
                
                \textit{``I think it's important (to be able to send private messages) if care is shared, especially between formal and informal caregivers."} [F-4]
                
                \textit{``There may be something for that particular caregiver that we don't want others to know or that it doesn't make much sense to share, it may even be sensitive information."} [F-5]
            \end{quotation}
       
            Caregivers, particularly the formal ones, also mentioned the importance of being notified of new messages (or observations) made about the \ac{PwD} to optimise the implementation of the sessions:
            \begin{quotation}
                \textit{``It's more pertinent in observations and messages, because that's what the caregiver is most involved in right away. It's what the caregiver needs to see right away because it may involve reorganizing things."} [F-2]
                
                \textit{``Even before starting the session the person should see if they have new messages because they may be relevant for the session.."} [F-5]
            \end{quotation}
       
            One of the formal caregivers also mentioned the importance of having messages (between formal and informal caregivers), as well as providing notification of them to informal caregivers, helping to reduce their isolation and also stimulating their memory (even though they may not have dementia):
            \begin{quotation}
                \textit{``I think this is very interesting (the messages) for several reasons, not only to share that information but it can serve to break the isolation, and that is another very interesting objective, to break the isolation. This possibility of sharing, of feeling that you are with others, right? To feel that others know what you are thinking and what you are doing can be very relevant... Feeling accompanied and feeling that others can accompany you too. That ends up being an almost univocal circuit. I feel accompanied and I feel that others can also accompany me better. It's in this perspective that I believe this breaks the isolation. Which is very relevant, especially nowadays..."} [F-7]
                
                \textit{``I think notifications are necessary when there is new information. In reminiscence therapy, we have to work on mnesic aspects, right? Thus, everything that helps with the visualization and to reinforcement (the memory) is very relevant, even for the informal caregiver... I think that there are always many benefits, we cannot forget informal caregivers may not have dementia but they may have memory problems due to several factors, namely the burnout that many experience."} [F-7]
            \end{quotation}
\section{Discussion}

\begin{figure*}[!h]
        \centering \includegraphics[width = 1\linewidth]{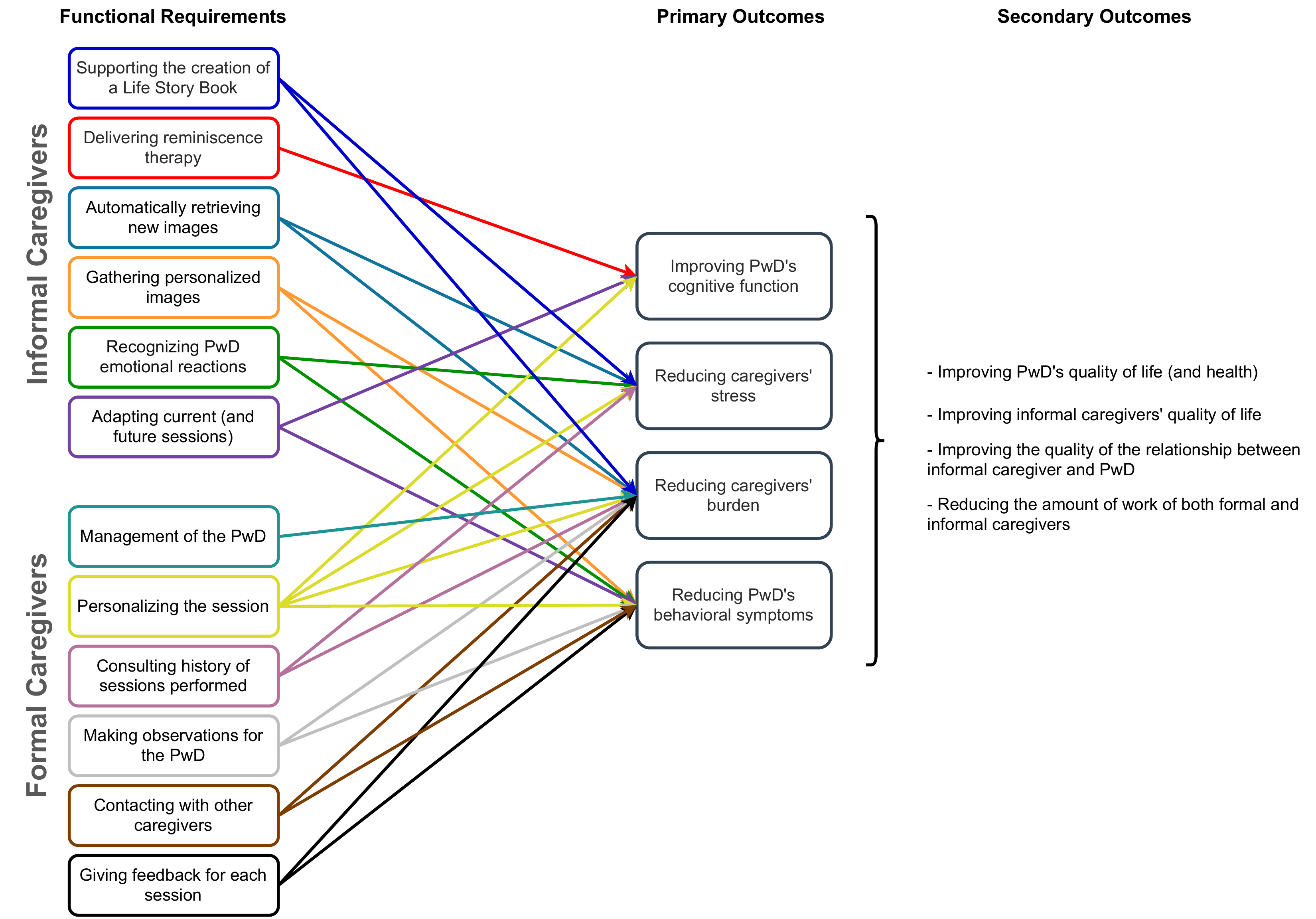}
    	 \caption{\label{fig:requirementsGatheringFinal} Summary of the final list of validated functional requirements  of technological solutions gathered for both formal and informal caregivers, as well as the corresponding expected primary and secondary outcomes. (figure best seen in color)}
    \end{figure*}

    As a result of our user-centered study, we present the final list of validated functional requirements to inform the development of technological solutions for both formal and informal caregivers (see Figure~\ref{fig:requirementsGatheringFinal}).
    It should be noted that some of the requirements outlined for one type of caregiver may also be useful for the other.
    We also present the expected primary and secondary outcomes from each requirement.

    The expected primary outcomes are the: improvement of the \ac{PwD} cognitive function, reduction of caregivers' stress and caregivers' burden, and reduction of the \ac{PwD} behavioral symptoms.
    As secondary outcomes, we foresee the improvement of the \ac{PwD}'s and caregivers quality of life (and health), improvement of the quality of the relationship between the informal caregiver and \ac{PwD}, and the decrease of the amount of work of the caregivers.
    
    The design of novel technological solutions within dementia care, particularly the ones targeted at reminiscing, should be the ability to \textit{Deliver reminiscence therapy} easily, since caregivers are subject to high levels of stress and overwhelmed with the amount of work they have (in particular, informal caregivers).
    
    Mechanisms to carry out the therapy in a simple way are a must have, although it should be also possible to manually program a session.
    Such mechanisms should help to create sessions fully (or semi-) automatically, reducing the amount of work caregivers would have to prepare and conduct the sessions, which will reduce their burden.
    In particular, for the informal caregivers, they will not have to prepare the session, thus they can perform it more often and focus in the memories and communication generated.
    We believe the quality of the relationship between the \ac{PwD} and the caregiver will increase.
    
    \newpage
    The use of an \ac{LSB} of the \ac{PwD} may prevent the abandonment of therapy due to lack of material (to be used in sessions), as well as to provide valuable information to formal caregivers to stimulate the \ac{PwD}'s memories.
    As such, technological solutions should focus in \textit{Supporting the creation of \ac{LSB}} to store multimedia related to the \ac{PwD} alongside with biographical information to be used in sessions:%
    \begin{quotation}
    	    \textit{``Creating a book of memories is not only useful for the \ac{PwD} but also for family members who may not always know very much about the person and the life they led many years ago. Family history is often lost because people wait too long to ask the right questions.''} [id-123]
    	    
    	    \textit{``Often people in my father in laws care home haven't got information to work with.''} [id-195]
    	    
            \textit{``My mother reads old letters with photos, but it would be nice to have it more organized. She also likes me to look up old friends on facebook and see if they have accounts.''} [id-269]
            
    	    \textit{``I love talking to my grandmother, she’s confident and can tell me intricate details of her life before Alzheimer’s. She forgets a 20 second old conversation but will happily discuss older events. It makes her less stressed knowing she remembers. I try to `visit' the past with her every time I'm with her, it’s like being in the room with another person..''} [id-364]
    	    
    	    \textit{``We use reminiscence as a tool for bonding during relative visits: our nephews and wives struggle to relate to her in her present condition, handing them an album to review stories leads her to interact.''} [id-370]
    	    
            \textit{``Memories of \ac{PwD} are difficult but with stimulation they take longer to forget. In everyday life, having tools is important when it comes to caring, if there isn't, you have to be creative as it differs for each \ac{PwD}.''} [id-456]
    \end{quotation}

    The \ac{LSB} might work as a starting point (for the technological solution) to create the initial sessions, as well as to diversify and personalize the following ones based on both emotional reactions and biographical information of the \ac{PwD}.
    For that, the technological solutions should be able to: i) \textit{Automatically retrieve new images}, in particular focusing on \textit{Gathering personalized images} (i.e., images related to moments of the life of the \ac{PwD}); ii) \textit{Recognizing \ac{PwD} emotional reactions}, and be able to \textit{Adapt the current (and future) sessions} based on that. 
    This is particularly important for the informal caregivers because, as we have seen, \ac{PwD} often become agitated, aggressive or angry, without the caregivers knowing how to properly deal with this situation, which ultimately impairs not only the emotional state and quality of life of both, as well as the quality of their relationship.
    Considering that the sessions' content would be more diversified and personalized, its quality is expected to increase as well.
    
    Following, we illustrate how difficult it is for caregivers to deal with \ac{PwD} negative emotional reactions, and how it affects their own well-being.
    \begin{quotation}
    	   \textit{``I can't - find it so difficult.''} [id-77]
    	   
    	   \textit{``I am persistent and try my best to not get agitated, which is very difficult..''} [id-91]
    	   
    	   \textit{``Go along with whatever she tell: truth or not! It's very hard, but not worth the melt down if you correct her!''} [id-101]
    	   
    	   \textit{``I yell because I get so frustrated.''} [id-148]
    	   
    	   \textit{``Talk. Sometimes I get upset to :(''} [id-178]
    	   
    	   \textit{``Try to calm her down but when she gets aggressive it is really difficult.''} [id-201]
    	   
    	   \textit{``I am unsure what to say.''} [id-272]
    	   
    	   \textit{``I try to keep her calm, change theme, and if she doesn't change after several intentions, I leave her alone, and come back after.''} [id-416]
    	   
    	   \textit{``It's stressful, frustrating, sad.''} [id-480]
    	   
    	   \textit{``Sometimes I don't know what to do.' }[id-534] 
    \end{quotation}
    
    All technological solutions targeted at reminiscence therapy should consider the multimedia as a crucial part of it, since without it, it is not possible to conduct the reminiscence therapy on a digital way.
    Personalized multimedia is well suited for the reminiscence therapy, and it should also be incorporated on all technological solutions regarding this therapy as generic content alone will not generate the response desired from the therapy.
    As such, being able to \textit{Personalize the session} is very important for the formal caregivers.
    They should be able to easily introduce new multimedia material, topics, and favorite items, as well as automatically adapt the session based on biographical information and the emotional reactions of the \ac{PwD}.
    Nonetheless, generic multimedia might be useful as well (e.g, formal caregivers are not always able to get personal content so they use generic ones).
    
    Formal caregivers should be able to \textit{Give feedback for each session}, and \textit{Consult the history of sessions performed}.
    We also believe that a functionality that allows to \textit{Make observations for the \ac{PwD}} should be provided as well.
    Such observations might be seen as a diary of the \ac{PwD} through the lens of the caregivers, which might be useful for future reminiscence therapy sessions but for other therapeutics as well.
    All the information collected (from the sessions performed) will help caregivers to compare sessions and improve them over time.
    The feedback registered will also be helpful to understand the overall emotional state of the \ac{PwD} alongside with their reactions to the stimuli presented.
    Ultimately, it will help reduce caregivers’ burden, while reducing \ac{PwD} behavioral symptoms.

    The \textit{Management of the \ac{PwD}} and share the care of a \ac{PwD} amongst multiple caregivers is a requirement appreciated and desired by the formal caregivers.
    The possibility to \textit{Contact with other caregivers} with whom they share the \ac{PwD} care should also be supported, allowing a centered management of the \ac{PwD} care.

    Finally, privacy is also very important: older people are not so used to new technologies, and they may be wary of technological solutions, in particular those in which they have to share their personal information.
    For that reason, and although it is not considered a functional requirement, such solutions should ensure the user that their information is secure and only those with permission can access it.
    Older people who are not familiar or comfortable with technology might need incentives to engage with these technologies and may have different usage patterns compared to younger people.
    Although it is outside the scope of the work presented, we draw attention to the importance of the design of novel technological solutions to make the workflow clear and easy to follow, and pay special attention to accessibility guidelines for the elderly (and their implications for design of user interfaces, if applicable).
    Moreover, we emphasize the need to provide help tools, since it is crucial for the success and adhesion to technological solutions.
    This is particularly meaningful for informal caregivers, since most of them are also elders, thus having more difficulties to learn new technologies.
    Failure to ensure the privacy of the information made available, or to provide help tools, can prevent people from accepting the technologies developed.

\section{Conclusions}

    In this work, we presented a user-centered study composed of three worldwide cross-sectional surveys, follow-up semi-structured interviews, and focus groups.
    Seven hundred and seven informal and 52 formal caregivers of \ac{PwD} participated in the study.
    
    Informal caregivers find themselves in distress and overburdened with the amount of work they have daily while caring for the \ac{PwD}. 
    Formal caregivers perform reminiscence therapy sessions more often than the informal ones.
    The number of informal caregivers performing the therapy has been decreasing, which we believe is related to the increase in caregivers' overload and stress due to the pandemic situation we are facing.
    In particular, we believe the existing lockdowns have conditioned the performance of therapy (and acquisition of material to be used in the sessions), increasing the isolation of the \ac{PwD}.
    Assistive technologies in dementia care may play an important role in aiding caregivers to deliver personalized and diversified therapy, reducing the amount of work they have to prepare and conduct the sessions.
    
    Our findings reveal that such technological solutions must provide mechanisms to carry out the therapy in a simple way, as well as diversify and personalize the current session (and following ones) based on both the biographical information of the \ac{PwD} and their emotional reactions (since \ac{PwD} often become agitated, aggressive or angry, and informal caregivers do not know how to deal with it).
    
    Formal caregivers need an easy way to manage \ac{PwD} and communicate with other caregivers, mechanisms to create therapy sessions, collect feedback after conducting each session, and consult the history of sessions performed (in particular, to identify images that triggered negative emotional reactions, and consult any notes taken about them).
    
    As a result of the study conducted, with the goal of inform the design of new assistive technological solutions, we close this work with a set of validated functional requirements gathered for the \ac{PwD} and both formal and informal caregivers, as well as the corresponding expected primary and secondary outcomes.

\section*{Funding}
    This work was supported by national funds through Fundação para a Ciência e a Tecnologia (FCT) through the LASIGE Research Unit, ref. UIDB/00408/2020 and ref. UIDP/00408/2020.
	Soraia M. Alarcão is funded by an FCT grant, ref. SFRH/BD/138263/2018.
	
\section*{CRediT authorship contribution statement}

    \textbf{Soraia M. Alarc\~{a}o}: Conceptualization, Methodology, Validation, Investigation, Formal analysis, Data curation, Writing - original draft. \textbf{Andr\'{e} Santana}: Methodology, Investigation, Formal analysis,  Data curation, Writing - review \& editing. \textbf{Carolina Maruta}: Conceptualization, Methodology, Validation, Writing - review \& editing. \textbf{Manuel J. Fonseca}: Conceptualization, Methodology, Validation, Writing - review \& editing, Supervision.

\section*{Declaration of Competing Interest}

    The authors declare that there are no known conflicts of interest associated with this publication and there has been no significant financial support for this work that could have influenced its outcome.

\section*{Acknowledgments}
	
    We would like to thank Filipa Brito for helping in the development and validation (of the first survey) from a psychological perspective; Fernando S\`{a}nchez, Laura Pereira, and V\^{a}nia Mendon\c{c}a for proofreading, respectively, the Spanish (first survey), English, and Portuguese versions of all surveys; finally, Isabella Mott for her help in the processing and coding of the data of the first survey.
    
    We would like to thank Irm\~{a}s Hospitaleiras - Casa de Sa\'{u}de da Idanha and Santa Casa da Miseric\'{o}rdia de Lisboa, the institutions involved in the focus groups, for their willingness to discuss and share their professional knowledge with us, allowing us to validate the requirements identified.
    
    Our special thanks goes to all the informal and formal caregivers worldwide for generously donating their time and sharing their knowledge with us, by participating in our study.

\bibliography{refs}

\end{document}